\documentclass[a4paper,10pt]{article}
\usepackage[utf8]{inputenc}
\sloppypar

\title{About a ``contextuality loophole'' in Bell's theorem claimed to exist by Nieuwenhuizen}
\author{I. Schmelzer}

\begin{document}

\maketitle

\begin{abstract}
Nieuwenhuizen argued that there exists some ``contextuality loophole'' in Bell's theorem. This claim is unjustified. In Bell's theorem non-contextuality is not presupposed but derived from Einstein causality using the EPR argument. 
\end{abstract}

In \cite{Nieuwenhuizen}, Nieuwenhuizen claims that there exists a ``contextuality loophole''.  In particular he claims:
\begin{quote}
\ldots even if Bell Inequality Violation is demonstrated beyond reasonable doubt, it will have no say on local realism, because absence of contextuality [sic] prevents the Bell inequalities to be derived from local realistic models.

\ldots Bell inequality violation (BIV) just proves that there cannot be a reduction to one common probability space. This finally implies that no conclusion can be drawn on local realism, since incompatible information can not be used to draw any conclusion. As explained below, the different pieces of the CHSH inequality involve fundamentally different distribution functions of the hidden variables, which cannot be put together in one over all covering distribution of all hidden variables of the set of considered experiments.
\end{quote}
This argument has to be rejected. Contextuality is, of course, important for understanding what really happens. But Bell was aware of this, and has known contextual hidden variable theories, in particular de Broglie-Bohm (dBB) theory.

Nieuwenhuizen proposes some ``improved hidden variables description'', where the formula
\[ C_{ij} = \int d\lambda \int d\lambda_{A_i} \int d \lambda_{B_j} \rho_{ij}(\lambda,\lambda_{A_i},\lambda_{B_j}) S_{A_i}(\lambda,\lambda_{A_i}) S_{B_j}(\lambda,\lambda_{B_j})\]
should be used to compute the result of the experiment. This formula would be compatible with dBB theory, with \(\lambda\) denoting the configuration of the state itself and \(\lambda_{A_i},\lambda_{B_j}\) describing the configuration of the measurement devices. So the idea that a contextual theory may allow to circumvent Bell's theorem is not new at all. It was well-known to Bell (a proponent of dBB theory) too. 

Now, in the famous formula (2) of Bell's paper \cite{theorem}
\begin{equation}
P(a,b) = \int d\lambda \rho(\lambda) A(a,\lambda)B(b,\lambda)
\end{equation}
is no room for contextuality. Instead, we have a single probability distribution \(\rho(\lambda)d\lambda\), and there is no dependence on hidden variables of the measurement devices. But does it follow that there is some ``contextuality loophole'' in Bell's theorem? Not at all.

The point is that the formula is not postulated (presupposing some non-contextuality) but derived in \cite{theorem}. The derivation is based on the EPR argument:
\begin{quote}
Now we make the hypothesis, and it seems one at least worth considering, that if the two measurements are made at places remote from one another the orientation of one magnet does not influence the result obtained with the other. Since we can predict in advance the result of measuring any chosen component of $\sigma_2$, by previously measuring the same component of $\sigma_1$, it follows that the result of any such measurement must actually be predetermined.                                                                                                          
\end{quote} 
It is this derived predetermination which closes the ``contextuality loophole'' for Einstein-causal and EPR-realistic theories. The predetermination of the result leaves  no place for hidden variables of the measurement devices, which could distort the results. One can formulate it differently: If the hidden variable of the measurement device would distort the measurement result at A, the information about the distorted measurement result would have to be transferred to B. Without such a transfer it would be impossible to obtain the same measurement result at B if Bob measures the same direction as Alice. But such a transfer of information would be forbidden by Einstein causality. 

The error made by Nieuwenhuizen remembers claims that determinism is presupposed in Bell's theorem. This error has been identified already by Bell himself in \cite{socks}:

\begin{quote}
It is important to note that to the limited degree to which determinism plays a role in the EPR argument, it is not assumed but inferred. What is held sacred is the principle of 'local causality' -- or 'no action at a distance'. \ldots  It is remarkably difficult to get this point across, that determinism is not a presupposition of the analysis.
\end{quote} 

Similarly, we can say: It is important to note that to the limited degree to which non-contextuality plays a role in Bell's theorem, it is not assumed but inferred. What is held sacred is the principle of 'local causality' -- or 'no action at a distance'. It is remarkably difficult to get this point across, that non-contextuality is not a presupposition of the analysis.


\begin{thebibliography}{99}
\bibitem{Nieuwenhuizen} T.M. Nieuwenhuizen, Is the Contextuality Loophole Fatal for the Derivation of Bell Inequalities? Found Phys 41: 580–591 (2011)

\bibitem{Bell} J.S.~Bell, Speakable and unspeakable in quantum mechanics, Cambridge University Press, Cambridge, 1987

\bibitem{theorem}
J.S.~Bell, On the Einstein-Podolsky-Rosen paradox, Physics 1, 195-200, 1964, also in \cite{Bell}, 14-21

\bibitem{socks} J.S.~Bell, Bertlmann's socks and the nature of reality, Journal de Physique, 42(3), C2 41-61 (1981), also in \cite{Bell}, 139-158

\end{thebibliography}
\end{document}